\documentclass[aps,prd,twocolumn,superscriptaddress,showpacs]{revtex4}

\usepackage{graphicx}

\usepackage{dcolumn}

\usepackage{bm}

\renewcommand\({\left(}
\renewcommand\){\right)}

\renewcommand\]{\right]}

\newcommand{\ra}{\rightarrow}

\def\lsim{\raise 0.4ex\hbox{$<$}\kern -0.8em\lower 0.62
ex\hbox{$\sim$}}

\def\gsim{\raise 0.4ex\hbox{$>$}\kern -0.7em\lower 0.62
ex\hbox{$\sim$}}

\def\lbar{{\hbox{$\lambda$}\kern -0.7em\raise 0.6ex
\hbox{$-$}}}

\newcommand\eq[1]{eq.~(\ref{#1})}
\newcommand\eqs[2]{eqs.~(\ref{#1}) and (\ref{#2})}
\newcommand\Eq[1]{Equation~(\ref{#1})}
\newcommand\Eqs[2]{Equations~(\ref{#1}) and (\ref{#2})}

\newcommand\pa{\partial}
\newcommand\p{\partial}

\newcommand\ee{\end{equation}}
\newcommand\be{\begin{equation}}
\def\bea{\begin{array}}
\def\eea{\end{array}}\def\ea{\end{array}}
\newcommand\ees{\end{eqnarray}}
\newcommand\bees{\begin{eqnarray}}

\def\p1{{\bf p}_1}
\def\p2{{\bf p}_2}
\def\k1{{\bf k}_1}
\def\k2{{\bf k}_2}

\newcommand{\dddM}{\kern 0.2em \raise 1.9ex\hbox{$...$}\kern -1.0em \hbox{$M$}}
\newcommand{\dddQ}{\kern 0.2em \raise 1.9ex\hbox{$...$}\kern -1.0em \hbox{$Q$}}
\newcommand{\dddI}{\kern 0.2em \raise 1.9ex\hbox{$...$}\kern -1.0em\hbox{$I$}}
\newcommand{\dddJ}{\kern 0.2em \raise 1.9ex\hbox{$...$}\kern-1.0em
\hbox{$J$}}
\newcommand{\dddcalJ}{\kern 0.2em \raise 1.9ex\hbox{$...$}\kern-1.0em
\hbox{${\cal J}$}}

\newcommand{\dddO}{\kern 0.2em \raise 1.9ex\hbox{$...$}\kern -1.0em
\hbox{${\cal O}$}}
\def\dddz{\raise 1.5ex\hbox{$...$}\kern -0.8em \hbox{$z$}}
\def\dddd{\raise 1.8ex\hbox{$...$}\kern -0.8em \hbox{$d$}}
\def\dddbd{\raise 1.8ex\hbox{$...$}\kern -0.8em \hbox{${\bf d}$}}
\def\ddbd{\raise 1.8ex\hbox{$..$}\kern -0.8em \hbox{${\bf d}$}}
\def\dddx{\raise 1.6ex\hbox{$...$}\kern -0.8em \hbox{$x$}}

\def\D{\Delta}
\def\p{\partial}

\def\nn{\nonumber}

\def\G{\Gamma}
\def\d{\delta}

\def\eps{\epsilon}

\def\dslash{\hspace{-1mm}\not{\hbox{\kern-2pt $\partial$}}}
\def\Dslash{\not{\hbox{\kern-4pt $D$}}}
\def\pslash{\not{\hbox{\kern-2.1pt $p$}}}
\def\kslash{\not{\hbox{\kern-2.3pt $k$}}}
\def\qslash{\not{\hbox{\kern-2.3pt $q$}}}

\newcommand{\inT}{\int_{-\infty}^{\infty}}

\newcommand{\Dl}{\int{\cal D}\lambda}

\begin{document}

\title{Path Integral Approach to non-Markovian First-Passage Time Problems}

\author{Michele Maggiore} 
\affiliation{D\'epartement de Physique Th\'eorique, 
Universit\'e de Gen\`eve, 24 quai Ansermet, CH-1211 Gen\`eve, Switzerland}
\author{Antonio Riotto}
\affiliation{
CERN, PH-TH Division, CH-1211, Gen\`eve 23,  Switzerland,
and INFN, Sezione di Padova, Via Marzolo 8,
I-35131 Padua, Italy}

\date{\today}

\begin{abstract}
The computation of the probability of
the first-passage time  through
a given threshold of a stochastic process is a classic problem that appears in
many branches of physics. 
When the stochastic dynamics is markovian, 
the probability admits elegant analytic solutions derived 
from the Fokker-Planck
equation with an absorbing boundary condition while, 
when the underlying dynamics is
non-markovian, the equation for the probability 
becomes non-local due to the appearance of
memory terms, and the problem becomes much harder to solve.
We show that  the
computation of the probability distribution and of the
first-passage time for
non-Markovian processes can be mapped into 
the evaluation of a path-integral with boundaries, and we develop a
technique for evaluating perturbatively this path integral,
order by order in the non-Markovian terms.

\end{abstract}





\maketitle

The computation of the statistical distribution of the  times at
which a stochastic process $\xi(t)$ first reaches a given threshold
(the so-called first-passage time problem) is a classic problem 
that appears in many
different contexts in physics, chemistry and biology. It is relevant 
for instance to
problems appearing in reaction rate theory,  nucleation theory or
neuron firing, to name just a few,
and it is treated  in a number of 
textbooks \cite{Stratonovich,vanKampen,redner2001} 
and reviews~\cite{Hangirmp}.
When the underlying dynamics is markovian, the function $\Pi(x, t)$
that gives the probability
distribution that $\xi(t)$ had a value $x$, in the continuum limit
satisfies a Fokker-Planck (FP) equation. The fact
that one is interested in the first-passage time problem
means that  one
wants to discard the trajectories once they have reached for the first
time a given threshold $x_c$. This is implemented imposing on the
FP equation an absorbing barrier boundary condition
$\Pi(x_c, t)=0$. The FP equation with this boundary condition,
together with the initial condition $\Pi(x, t=0)=\d_D(x=0)$, where
$\d_D$ is the Dirac delta, can be
elegantly solved using the method of images, and one finds \cite{Chandra}
\be\label{Pix0}
\Pi(x;t)
= \frac{1}{\sqrt{2\pi t}}\,
\[e^{-x^2/(2t)}-
e^{-(2x_c-x)^2/(2t)}\]\, .
\ee
When the underlying dynamics is
non-Markovian, however, the problem becomes much more difficult.
The system 
acquires memory properties and 
the probability $\Pi(x, t)$ no longer satisfy a simple diffusion
equation such as the 
FP equation. 
Furthermore,   the correctness of the 
``absorbing barrier'' boundary condition 
is now far from obvious  \citep{vKampen,knessl}.
For these reasons,
first-passage  problems for non-Markovian processes
are known to be very hard to solve, and have been attacked 
in various way, see e.g. 
\citep{hanggi1981,weiss1983,west,Sire,vkampen1998,Vere} and references therein.
Our original interest in the problem arose from a specific  question in
cosmology, namely the computation of the mass distribution of dark
matter halos generated by the evolution
of non-Gaussian primordial density fluctuations, 
which can indeed be formulated as a first-passage time
problem with non-Markovian dynamics~\cite{Bond}. We think however that
the techniques that we have developed in 
refs.~\cite{MR1,MR2,MR3}, and which allowed us to solve our problem, 
have a broader interest, and we find it useful to
present them here in a more general context. 

Let $\xi(t)$ be a variable that evolves stochastically with time
$t$, with  $\langle\xi(t)\rangle =0$. We consider an ensemble of
trajectories starting at $t_0=0$ from an initial position
$\xi(0) =x_0$, and we follow them for a
time $t$.  
We discretize the interval $[0,t]$ in steps
$\D t=\eps$,  so $t_k=k\eps$ with $k=1,\ldots n$.
A trajectory is then defined by
the collection of values $\{x_1,\ldots ,x_n\}$, such that $\xi(t_k)=x_k$.
There is no absorbing barrier, i.e. $\xi(t)$ is allowed to range freely from
$-\infty$ to $+\infty$.
The probability density in the space of  trajectories is 
\be\label{defW}
W(x_0;x_1,\ldots ,x_n;t_n)= \langle \prod_{i=1}^{n}
\d_D (\xi(t_i)-x_i)\rangle\, .
\ee
In terms of $W$ we  define
\be\label{defPi}
\Pi_{\eps} (x_0;x_n;t_n)
 \equiv
\prod_{i=1}^{n-1}\int_{-\infty}^{x_c} dx_i\,
W(x_0;x_1,\ldots ,x_n;t_n)\, ,
\ee
where  $t_n=n\eps\equiv t$ and we will often write $x_n$ simply as $x$.
So, 
$\Pi_{\eps} (x_0;x;t)$ is the probability density of arriving in
$x$ at time $t$,   starting from $x_0$ at time $t_0=0$, 
through trajectories that never exceeded $x_c$.
Observe that the final point
$x$  ranges over $-\infty<x<\infty$. For later use, we
find useful to write explicitly 
that  $\Pi$ depends also on the temporal
discretization step $\eps$.
We are finally interested in its continuum limit, $\Pi_{\eps=0}$, and
we will see in due course that taking the limit $\eps\ra 0$ 
of $\Pi_{\eps}$ is non-trivial.

The usefulness of $\Pi_{\eps}$ is that it allows us to compute the
first-crossing rate from first principles, without the need of
postulating the existence of an absorbing barrier. Simply,
the quantity
$\int_{-\infty}^{x_c}dx\, \Pi_{\eps}(x_0;x;t)$
gives the probability that at time $t$ a trajectory always stayed in
the region $x<x_c$, for all times smaller than $t$. The rate of change
of this quantity is therefore equal to minus the rate at which trajectories
cross for the first time the barrier, so 
the first-crossing rate is
\be\label{prFirst}
{\cal F}(t)=-\int_{-\infty}^{x_c}dx_n\,\pa_t \Pi_{\eps}(x_0;x_n;t)\, .
\ee
Observe that no reference to a hypothetical ``absorbing barrier'' is made in
this formalism.  We will see below how  an effective
absorbing barrier emerges from this microscopic approach.

The probability density $W$ can be expressed in terms of
the  connected 
correlators $\langle \xi_{i_1}\ldots\xi_{i_p}\rangle_c$
as~\cite{Stratonovich}
\bees\label{WnNG}
&&W(x_0;x_1,\ldots ,x_n;t_n)=\Dl \,\,  e^{i\sum_{i=1}^n\lambda_ix_i}
\\
&&\times \exp\{ \sum_{p=2}^{\infty} \frac{(-i)^p}{p!}\,
 \sum_{i_1=1}^n\ldots \sum_{i_p=1}^n
\lambda_{i_1}\ldots\lambda_{i_p}\,
\langle \xi_{i_1}\ldots\xi_{i_p}\rangle_c\}\, ,\nn
\ees
where
$\Dl \equiv
\inT\frac{d\lambda_1}{2\pi}\ldots\frac{d\lambda_n}{2\pi}$
and $\xi_i=\xi(t_i)$. The problem is therefore reduced to computing the
path-integral (\ref{defPi}), over variables $x_i$ bounded by $x_c$,
with $W$ given by \eq{WnNG}.

We first consider
the simple case in which $\xi$ has
gaussian statistics (so only the two-point connected function is
non-vanishing), and obeys 
a Langevin equation
$\dot{\xi}=\eta(t)$
with
a noise $\eta$ whose correlator is a Dirac delta,
$\langle \eta(t)\eta(t')\rangle =\d_D (t-t')$.
In this case
the 2-point correlator
is easily computed
\be
\langle\xi(t_i)\xi(t_j)\rangle_c=\int_0^{t_i}dt\int_0^{t_j}dt'
\langle\eta(t)\eta(t')\rangle
={\rm min}(t_i,t_j)\, ,
\ee
and from this, performing the gaussian integrals in \eq{WnNG}, we find
\be\label{W}
W^{\rm gau}(x_0;x_1,\ldots ,x_n;t_n)=\frac{1}{(2\pi\eps)^{n/2}}\, 
e^{-\frac{1}{2\eps}\sum_{i=0}^{n-1}  (x_{i+1}-x_i)^2}\, .
\ee
where we denote by $W^{\rm gau}$ the value of $W$ when $\xi$ has 
gaussian statistics and obeys a Langevin equation with Dirac-delta
noise. 
Not surprisingly, in this case  we got
the Wiener measure.
To compute $\Pi^{\rm gau}_{\eps}$
by performing directly the integrals over 
$x_1, \ldots ,x_{n-1}$ in \eq{defPi},
and then taking the limit $\eps\ra 0$
is very difficult, 
since the integrals in \eq{defPi}
run only up to $x_c$, and already the inner
integral gives an error function whose argument
involves the next integration variable.
In \cite{MR1} we have then followed a different route.
Using the explicit form (\ref{W}) we proved that
\be\label{CK3}
\Pi^{\rm gau}_{\eps} (x_0;x;t+\eps)=
\int_{x-x_c}^{\infty}\hspace{-2mm} d(\D x)\, \Psi_{\eps}(\D x)
\Pi^{\rm gau}_{\eps} (x_0;x-\D x;t),
\ee
where
$\Psi_{\eps}(\D x)=(2\pi\eps)^{-1/2} 
\exp\{-(\D x)^2/(2\eps) \}$.
\Eq{CK3} generalizes the
Chapman-Kolmogorov equation, to which it reduces if we send
$x_c\ra\infty$, i.e. if the integrations
in  \eq{defPi} are not bounded, and
expresses the fact that the evolution corresponding to a
Langevin equation with Dirac-delta noise is
a markovian process. 
\Eq{CK3} allows us to compute the continuum limit
of $\Pi^{\rm gau}_{\eps}$, as follows.
In the limit $\eps\ra 0$ we have
$\Psi_{\eps}(\D x)\ra \d_D(\D x)$. 
If $x-x_c<0$, the integral in \eq{CK3}
includes the support of the Dirac delta, 
and we just get
the trivial identity that
$\Pi^{\rm gau}_{\eps=0} (x_0;x;t)$ is equal to itself. However, 
if $x-x_c> 0$, the
right-hand side vanishes and we get
$\Pi^{\rm gau}_{\eps=0} (x_0;x;t)=0$. 
Therefore we
find that
$\Pi^{\rm gau}_{\eps=0} (x_0;x;t)=0$ if $x\geq x_c$. 
(If $x=x_c$ only 
one half of the
support of $\Psi_{\eps}$ is inside the integration region,  so we get
$\Pi^{\rm gau}_{\eps=0} (x_0;x_c;t)=(1/2)\Pi^{\rm gau}_{\eps=0} (x_0;x_c;t)$, 
which again 
implies $\Pi^{\rm gau}_{\eps=0} (x_0;x_c;t)=0$).
This is the
boundary condition that in the usual treatment is just imposed by
hand, while here it follows from the formalism.
Consider now \eq{CK3} when $x<x_c$.  In this case the zeroth-order term  in
$\eps$ gives
a trivial identity. 
Pursuing the expansion to higher orders one finds that, in the limit
$(x_c-x)/\sqrt{\eps}\ra 0^+$, and therefore when $x$ is
fixed and strictly smaller
than 
$x_c$ while $\eps\ra 0^+$, the dependence on the index $\eps$ in
$\Pi^{\rm gau}_{\eps}$ can be expanded in integer powers of $\eps$,
\be\label{CK8}
\Pi^{\rm gau}_{\eps} (x_0;x;t) =\Pi^{\rm gau}_{\eps=0} (x_0;x;t)
+\eps \Pi^{\rm gau}_{(1)} (x_0;x;t) 
+\ldots\, .
\ee
Collecting terms of the same order
in $\eps$  we then find that, for $x<x_c$,
$\Pi^{\rm gau}_{\eps=0}(x_0;x;t)$
satisfies a FP equation.
We therefore end up with a FP equation with
the boundary condition $\Pi^{\rm gau}_{\eps=0}(x_0;x=x_c;t)=0$, so
we recover \eq{Pix0}. We have therefore succeeded in deriving this
standard result
from our path integral approach. Observe  that the
boundary condition $\Pi^{\rm gau}_{\eps=0}(x_0;x=x_c;t)=0$ emerges
only when we take the continuum limit, and does not hold for finite $\eps$.

Having
computed the path-integral in the
markovian case, we can tackle the problem of non-Markovian
dynamics treating the non-Markovian terms as perturbations.
For illustration, we discuss the case of colored gaussian
noise, i.e. again the only non-vanishing connected correlator is the
two-point correlator, but now we take it to have the form
$\langle\xi(t_i)\xi(t_j)\rangle_c
={\rm min}(t_i,t_j) +\D (t_i,t_j)$,
for some function $\D (t_i,t_j)\equiv \D_{ij}$. So we want to compute
\bees
&&\Pi_{\eps}(x_n;t_n) =
\int_{-\infty}^{x_c} 
dx_1\ldots dx_{n-1}\,\Dl\nn\\
&&\hspace*{-2mm}\times\exp\left\{
i\lambda_ix_i -\frac{1}{2}
[{\rm min}(t_i,t_j) + \D(t_i,t_j)]
\lambda_i\lambda_j\right\} ,\label{Delta1}
\ees
where, for simplicity, we set $x_0=0$ and we eliminated it from the
list of variables on which $\Pi_{\eps}$ depends.
We assume
that $\D_{ij}$ is proportional to a small parameter, and we 
expand perturbatively in $\D_{ij}$.
Using
$\lambda_ke^{i\lambda x}=-i\pa_xe^{i\lambda x}$ and writing
$\pa/\pa x_i=\pa_i$, the first-order
correction
to
$\Pi_{\eps}$, that we denote by $\Pi^{{\D}1}_{\eps}$, is
\be\label{Delta3}
\Pi^{{\D}1}_{\eps}(x_n;t_n) 
=\sum_{i,j=1}^n \frac{\D_{ij}}{2}\int_{-\infty}^{x_c} 
dx_1\ldots dx_{n-1}\,\pa_i\pa_j
W^{\rm gau}
\, .
\ee
We rewrite the term $ \D_{ij}\pa_i\pa_j$ separating
explicitly the derivative $\pa_n\equiv \pa/\pa x_n$
from the derivatives $\pa_i$ with
$i<n$.  
Let us at first consider the case in which $\D(t_i,t_j)$ vanishes at
least as $t_j-t_i$ as $t_i\ra t_j$. This was indeed the case in the
application to the cosmological problem discussed in \cite{MR1}. The
more general case will be discussed later.
In this case, using also $\D_{ij}=\D_{ji}$,
\be\label{prsumD}
\frac{1}{2} 
\sum_{i,j=1}^n \D_{ij}\pa_i\pa_j
=\sum_{i=1}^{n-1}\D_{in}\pa_i\pa_n+
\sum_{i<j}\D_{ij}\pa_i\pa_j\, ,\nn
\ee
where 
$\sum_{i<j}\equiv \sum_{i=1}^{n-2}\sum_{j=i+1}^{n-1}$.
When inserted into \eq{Delta3}
the term
$\sum_{i=1}^{n-1}\D_{in}\pa_i\pa_n$ brings a factor $\sum_i$ that, in
the continuum limit, produces an integral over an intermediate
time $t_i$. Because of this dependence on the past history, we call
this the ``memory term''. Similarly, the double sum in
$\sum_{i<j}\D_{ij}\pa_i\pa_j$ gives, in the
continuum limit, a double integral over intermediate times $t_i$ and $t_j$,
and we call it the ``memory-of-memory'' term.  Thus,
$\Pi^{{\D}1}_{\eps}=
\Pi_{\eps}^{\rm mem}+\Pi_{\eps}^{\rm mem-mem}$,
where
\be\label{defPimemory}
\Pi_{\eps}^{\rm mem}
=\sum_{i=1}^{n-1}\D_{in}\pa_n
\int_{-\infty}^{x_c} 
dx_1\ldots dx_{n-1}\,\pa_iW^{\rm gau}\, ,
\ee
\be\label{defPimem-memory}
\Pi_{\eps}^{\rm mem-mem}=\sum_{i<j}\D_{ij}
\int_{-\infty}^{x_c} 
dx_1\ldots dx_{n-1} \pa_i\pa_j\,W^{\rm gau}
\, .
\ee
To compute the memory term we  integrate $\pa_i$ by parts
and we make use of the fact that $W^{\rm gau}$ satisfies
\bees\label{facto}
&&W^{\rm gau}(x_0;x_1,\ldots, x_{i}= x_c, \ldots ,x_n; t_n)\\
&=& 
W^{\rm gau}(x_0;x_1,\ldots, x_{i}= x_c;t_i)\nn\\
&&\times W^{\rm gau}(x_c; x_{i+1}, \ldots ,x_n; t_n-t_i)\, ,\nn
\ees
as can be checked from the explicit expression (\ref{W}), so
\bees
&&\int_{-\infty}^{x_c} 
dx_1\ldots dx_{n-1} \pa_i\,W^{\rm gau}\nn\\
&=& \Pi^{\rm gau}_{\eps}(x_0;x_c;t_i)
\Pi^{\rm gau}_{\eps}(x_c;x_n;t_n-t_i)\, .
\ees
In the continuum limit (if the integral converges, as we will check in
a moment), we  replace
$\sum_{i=1}^{n-1}$ by $(1/{\eps})\int_0^{t_n} dt_i$
and we get
\bees
&&\Pi_{\eps=0}^{\rm mem}(x_n;t_n)= \pa_n \int_0^{t_n} dt_i\,
\D(t_i,t_n)\label{pr1}\\
&&\times \lim_{\eps\ra 0^+} 
\frac{1}{\eps} \Pi^{\rm gau}_{\eps}(x_0;x_c;t_i)
\Pi^{\rm gau}_{\eps}(x_c;x_n;t_n-t_i)\, .\nn
\ees
It is quite interesting to observe that the memory term is determined
by the finite-$\eps$ corrections to the markovian term. We found 
above that
$\Pi^{\rm gau}_{\eps =0}(x_0;x_n;t)$ vanishes for $x_n=x_c$. 
However, we see from 
\eq{pr1} that it is not enough to know that 
$\Pi^{\rm gau}_{\eps}(x_0;x_c;t)=0$ for $\eps\ra 0$, but we also need
to know how fast it goes to zero with $\eps$. For
$x-x_c$ fixed and strictly negative, in the limit $\eps\ra 0$ we have seen
above that the correction to $\Pi^{\rm gau}_{\eps=0}$ 
are ${\cal O}(\eps)$. However, in \eq{pr1} we need 
$\Pi^{\rm gau}_{\eps}$ for $x=x_c$. In this case the form of the
correction changes qualitatively.
Technically this comes from the fact that,
after changing the integration variables from $x_i$ to
$y_i=x_i/\sqrt{2\eps}$, which makes the exponential factors in 
\eq{W} independent of $\eps$, the lower integration limit
in \eq{CK3} becomes $(x-x_c)/\sqrt{2\eps}$. For 
$x$ fixed and strictly smaller that $x_c$, when $\eps\ra 0^+$ this
lower limit goes to $-\infty$, while for $x=x_c$ it is zero for all
$\eps$, resulting in a different form of the solution. The passage
between the two regimes takes place when the lower limit of the
integral is ${\cal O}(1)$, i.e. when $x-x_c\sim \sqrt{\eps}$.
So, in the continuum limit, the
path integral in \eq{defPi} has three different regimes: for $x$
strictly smaller than $x_c$, with $x_c-x$ finite, it approaches
\eq{Pix0}, plus corrections ${\cal O}(\eps)$. For $x>x_c$ and
$x-x_c$ finite it is
equal to zero, plus corrections which can be shown to be 
exponentially small in $\eps$, 
${\cal  O}\(\exp\{-(x_c-x)^2/(2\eps)\}\)$. These two regimes are
connected by an infinitesimal boundary layer, of thickness
$|x-x_c|={\cal O}(\sqrt{\eps})$, where the corrections to the
continuum solution are themselves
${\cal O}(\sqrt{\eps})$. In particular, when 
$x=x_c$, we find in \cite{MR1} that, for $x_0<x_c$,
\be\label{Pigammafinal}
\Pi^{\rm gau}_{\eps} (x_0;x_c;t)=
\sqrt{\eps}\, 
\frac{x_c-x_0}{\sqrt{\pi}\, t^{3/2}} e^{-(x_c-x_0)^2/(2t)} +{\cal O}(\eps)\, .
\ee
We see that the factor $\sqrt{\eps}$ from
$\Pi^{\rm gau}_{\eps}(x_0;x_c;t_i)$ and a similar
factor $\sqrt{\eps}$ from
$\Pi^{\rm gau}_{\eps}(x_c;x_n;t_n-t_i)$, cancel the factor $1/\eps$ which
comes from transforming the sum into an integral. The limit in
\eq{pr1} is then finite, and
\bees\label{Pimem2}
\Pi_{\eps=0}^{\rm mem}(x_n;t_n)&=&\frac{1}{\pi}
\pa_n\int_0^{t_n}dt_i\,
\D(t_i,t_n)\frac{x_c(x_c-x_n)}{t_i^{3/2}(t_n-t_i)^{3/2}}
\nn\\
&&\times
\exp\left\{-\frac{x_c^2}{2t_i}-\frac{(x_c-x_n)^2}{2(t_n-t_i)}
\right\}\, .
\ees
Recall that $\Pi^{\rm gau}_{\eps}(x_0;x_c;t_i)$ represents the probability
density for trajectories that start at $x_0$ at the initial time and
arrive at $x_c$ at time $t_i$, staying always in the region $x\leq
x_c$,
for a variable obeying gaussian statistics and
driven by a Dirac delta noise. 
Thus, \eq{pr1}
has a vivid diagrammatic interpretation in terms of a sum over the
markovian 
trajectories that start at $x_0$, touch for the first time
the boundary $x_c$ at an
intermediate time $t_i$, but rather than crossing the threshold
go back into the region $x<x_c$, finally reaching a value $x_n$
at time $t_n$. We see that the probability density
$\Pi^{\rm gau}_{\eps}(x_0;x_c;t_i)$ plays the role 
that in the perturbative expansion of the path
integral in quantum field theory is played by the free propagator, and
the whole complexity of non-Markovian dynamics enters through the
presence of the boundary at $x=x_c$.

The memory-of-memory term can be computed in the same way. Now we
integrate by parts the two derivatives $\pa_i\pa_j$ in
\eq{defPimem-memory}. This leaves us with
$W$ evaluated in $x_i=x_c$ and $x_j=x_c$. Using \eq{facto} we write it
as a product of three terms, and
\bees
&&\Pi_{\eps}^{\rm mem-mem}(x_0;x_n;t_n)=\sum_{i<j}\D_{ij}
\Pi^{\rm gau}_{\eps}(x_0;x_c;t_i)\nn\\
&&\times
\Pi^{\rm gau}_{\eps}(x_c;x_c;t_j-t_i)
\Pi^{\rm gau}_{\eps}(x_c;x_n;t_n-t_j)\, .\label{Pimemmem1}
\ees
Diagrammatically, this corresponds to a sum over the  markovian 
trajectories that start at $x_0$, touch  for the first time
the boundary $x_c$ at an intermediate time $t_i$,
go back into the region $x<x_c$, touch again the boundary at time
$t_j$, and
finally reach the value $x_n$
at time $t_n$,   always staying in the region $x_i\leq x_c$ for $i<n$.
The explicit computation shows that
$\Pi^{\rm gau}_{\eps} (x_c;x_c;t)=\eps/(\sqrt{2\pi}\, t^{3/2})$.
Then again the continuum limit is finite, and 
\bees
&&\Pi_{\eps=0}^{\rm mem-mem}(x_0=0;x_n;t_n)=
\frac{1}{\pi\sqrt{2\pi}}\, x_c(x_c-x_n)\nn\\
&&\times
\int_0^{t_n}dt_i\, \int_{t_i}^{t_n}dt_j
\frac{\D(t_i,t_j)}{t_i^{3/2}(t_j-t_i)^{3/2} (t_n-t_j)^{3/2}}\nn\\
&&\times
\exp\left\{-\frac{x_c^2}{2 t_i}
-\frac{(x_c-x_n)^2}{2(t_n-t_j)}\right\}\, .\label{prmemmem}
\ees
\Eqs{Pimem2}{prmemmem} provides an analytic expression for the
first-order non-Markovian corrections due to a non-trivial two-point
function, under the assumption that
$\D(t_i,t_j)$ vanishes, at least linearly, as
$t_i\ra t_j$. Given the probability distribution, 
the first-crossing rate is given by \eq{prFirst}, and from this one
can derive
all the properties
concerning the statistics of the time at which the threshold is first
crossed. 

\begin{figure}
\includegraphics[width=0.45\textwidth]{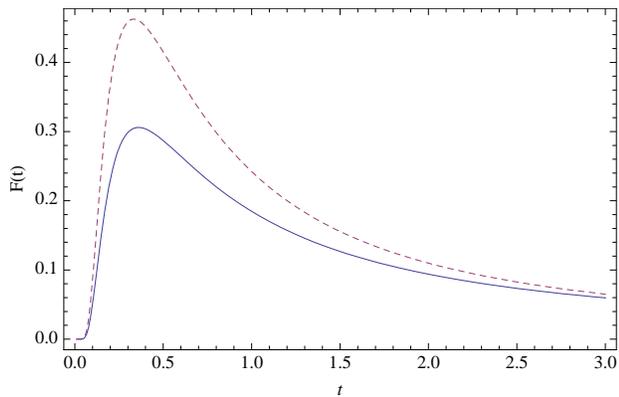}
\caption{\label{figFt}
The first crossing rate ${\cal F}(t)$ from \eq{Ffinal}
for the non-Markovian case with $\kappa=0.44$ (blue solid line)
compared to the markovian result obtained setting $\kappa=0$.
In both cases we set  $x_c=1$.
}
\end{figure}

As an application, we illustrate our results setting
$\D_{ij}=\kappa t_i(t_j-t_i)/t_j$ for $t_i\leq t_j$. (The value for
$t_i\geq t_j$ is obtained by symmetry, $\D_{ij}=\D_{ji}$).
This was indeed the form of $\D_{ij}$ in the problem studied in \cite{MR1},
where
$\kappa\simeq 0.44$ played
the role of the  expansion parameter. All integrals can then be
computed analytically, and for the first-crossing rate we find
\be\label{Ffinal}
{\cal F}(t)=\frac{1-\kappa}{\sqrt{2\pi}}\, 
\frac{x_c}{t^{3/2}} e^{-x_c^2/(2t)}+
\frac{\kappa}{2\sqrt{2\pi}} 
\frac{x_c}{t^{3/2}}\G\(0,\frac{x_c^2}{2t}\)\, ,
\ee
where $\Gamma (0,z)$ is the incomplete Gamma function. The result is
shown in Fig.~\ref{figFt}, together with the markovian case, which is
obtained setting
$\kappa =0$.

In the above example, $\D(t_i,t_j)$ goes to zero linearly as
$t_j-t_i\ra 0$.
If however $\D(t_i,t_j)$ goes to a 
non-zero constant at $t_i=t_j$, we see
that the integral over $dt_j$ in \eq{prmemmem} diverges at the lower
limit $t_j=t_i$. At the same time, in \eq{prsumD} we must also include
a term proportional to $\D_{ii}\pa_i^2$, since $\D_{ii}$ is
non-zero, and this term also leads to a divergent integral. 
After  regularizing the sums over $i,j$ one can
extract the divergent part of both integrals, which are both
proportional to $1/\sqrt{\eps}$, plus the finite part. The divergent
parts of these two terms cancel among them, and we remain with a
finite result.  We refer the reader to 
appendix~B of ref.~\cite{MR1} for details.

The same strategy can be applied to all higher-order correlators.
In ref.~\cite{MR3} we performed the computation expanding
\eqs{defPi}{WnNG} to linear order
in the three-point
correlator $\langle \xi_i\xi_j\xi_k\rangle$.  The result fits quite well the
outcome of cosmological $N$-body simulations with non-Gaussian
initial conditions \cite{Grossi}, giving further confidence
in our technique.

In conclusion,  we have developed a  very general method for
computing systematically the non-Markovian contributions to the
first-crossing rate, 
whenever they can be treated as perturbations of the
markovian dynamics. 

\vspace{5mm}

\noindent We thank Michel Droz, 
Sabino Matarrese and Sidney Redner
for useful discussions. 
The work
of MM is supported by the Fond National Suisse. 
The work of AR is supported by 
the European Community's Research Training Networks 
under contract MRTN-CT-2006-035505.


\begin{thebibliography}{99}

\bibitem{Stratonovich} 
R. L. Stratonovich,  
``Topics in the Theory of
  Random Noise'',  Gordon and Breach, New York, 1967.

\bibitem{vanKampen} N. G. van Kampen,   ``Stochastic Processes in
  Physics and Chemistry'', North-Holland, Amsterdam 1992.

\bibitem{redner2001}
S. Redner,  
``A guide to first-passage processes", Cambridge University
Press, 2001.

\bibitem{Hangirmp} P. H\"{a}nggi, P. Talkner and M.~Borkovec,
Rev. Mod. Phys. {\bf 62},  251 (1990).

\bibitem{Chandra} 
S. Chandrasekhar,  Rev. Mod. Phys. {\bf 15},  1 (1943).

\bibitem{vKampen} 
N. G. van Kampen  and  I. Oppenheim,  
J. Math. Phys. {\bf 13}  842 (1972).

\bibitem{knessl} 
C. Knessl, {et~al.}  J. Stat. Phys. {\bf 42},  169 (1986).

\bibitem{hanggi1981}
P. H\"{a}nggi, Z. Phys. {\bf B45}, 79 (1981).

\bibitem{weiss1983}
G. H. Weiss   et al.,   Physica {\bf 119A}, 569 (1983).

\bibitem{west} J. Masoliver, K.~Lindenberg and B.~J.~West,
Phys. Rev. {\bf A33} (1986), 2177.

\bibitem{Sire} S. N. Majumdar and C.~Sire,
Phys. Rev. Lett. {\bf 77}, 1420 (1996).


\bibitem{vkampen1998}
N. G. van Kampen, Braz. Journ. of Phys. {\bf 28}, 90 (1998).

\bibitem{Vere} T. Verechtchaguina, I. M. Sokolov and
  L.~Schimansky-Geier, Europhys. Lett. {\bf 73}, 691 (2006).

\bibitem{Bond}
J.~R.  Bond, {et~al.}  
  Astrophys. Journal.  {\bf 379},  440 (1991).

\bibitem{MR1} Maggiore, M. \& A. Riotto, arXiv:0903.1249 [astro-ph].

\bibitem{MR2} Maggiore, M. \& A. Riotto, arXiv:0903.1250 [astro-ph].

\bibitem{MR3} Maggiore, M. \& A. Riotto, arXiv:0903.1251 [astro-ph].

 
\bibitem{Grossi}
M. Grossi et al., 
arXiv:0902.2013 [astro-ph].


\end{thebibliography}
\end{document}